# PCPT and ACPT: Copyright Protection and Traceability Scheme for DNN Models

Xuefeng Fan[1], Dahao Fu[2], Hangyu Gui[2], Xinpeng Zhang[3] and Xiaoyi Zhou[2]
[1]School of Information and Communication Engineering, Hainan University, Haikou, China
[2]School of Cyberspace Security, Hainan University, Haikou, China
[3]School of Communication and Information Engineering, Shanghai University, Shanghai, China

*Abstract*—Deep neural networks (DNNs) have achieved tremendous success in artificial intelligence (AI) fields. However, DNN models can be easily illegally copied, redistributed, or abused by criminals, seriously damaging the interests of model inventors. The copyright protection of DNN models by neural network watermarking has been studied, but the establishment of a traceability mechanism for determining the authorized users of a leaked model is a new problem driven by the demand for AI services. Because the existing traceability mechanisms are used for models without watermarks, a small number of false-positives are generated. Existing black-box active protection schemes have loose authorization control and are vulnerable to forgery attacks. Therefore, based on the idea of black-box neural network watermarking with the video framing and image perceptual hash algorithm, a passive copyright protection and traceability framework *PCPT* is proposed that uses an additional class of DNN models, improving the existing traceability mechanism that yields a small number of false-positives. Based on an authorization control strategy and image perceptual hash algorithm, a DNN model active copyright protection and traceability framework *ACPT* is proposed. This framework uses the authorization control center constructed by the detector and verifier. This approach realizes stricter authorization control, which establishes a strong connection between users and model owners, improves the framework security, and supports traceability verification.

*Index Terms*—DNN, PCPT, ACPT, traceability, copyright protection.

## I. INTRODUCTION

Deep neural networks (DNNs) have been widely used in image and speech processing, including natural language processing, computer vision, image processing, and speech recognition. Advanced neural network models, including LeNet, VGGNet, GoogLeNet, ResNet, and DenseNet, have also shown excellent performance from generic object recognition to object detection and face recognition[1]. Internet companies, such as Microsoft, Baidu, and Google, have deployed DNN models in their products and services to provide intelligent and high-quality services. In contrast to traditional multimedia data, the cost of training a good DNN model is considerable. It requires the use of large-scale datasets, huge computing resources, and large labor costs. Therefore, the copyright protection and traceability of DNN models are particularly important. The methods to protect the copyright of DNN models from being illegally stolen, plagiarized and to trace the source of the stolen model to determine the authorized user of that leaked model are both challenging problems.

On the one hand, inspired by traditional watermarking ideas [2, 3], researchers have proposed various schemes for the copyright protection of DNN models [4], which can be roughly divided into four categories: white-box watermarking [5, 6], black-box watermarking [7, 8], gray-box watermarking [9], and null-box watermarking [10, 11]. However, according to the literature, only the KeyNet framework proposed by Jebreel *et al.* [12] has addressed the problem of traceability after the DNN model is illegally stolen and distributed. However, when the KeyNet framework is used for models without watermarks, a small number of false-positives are produced. For example, the VGG16 and ResNet18 networks yield 7.92% and 18.92% false-positive rates, respectively. Therefore, based on the idea of black-box neural network watermarking and the video framing and image perceptual hash algorithm, a passive copyright protection and traceability (PCPT) framework for DNN models using additional classes is proposed. Since PCPT uses additional classes as the trigger set, the distortion of the original decision boundary is minimized (or even eliminated), thus realizing a zero false-positive rate in the unlabeled model. Specifically, the PCPT framework utilizes video material shot by the model owner as a key. After a video is framed, several trigger sets are constructed according to the different subjects in the video. When the owner information is embedded in the trigger set using the image perceptual hash algorithm, the set is used as an additional class, and different additional classes are embedded into DNN models as watermarks with fine-tuning technology. The model is then distributed to different users to realize traceability. Note that one DNN watermark can be used for model copyright protection, while different watermarks embedded in different DNN model copies can be used for model traceability.

On the other hand, the demand for emerging industries such as smart homes and self-driving cars is forcing researchers and developers to adopt DNNs to address intelligent interaction and control. In these scenarios, the underlying embedded systems typically run DNNs locally to address latency and privacy issues [13]. It requires a service similar to an Software Development Kit (SDK). Proprietary SDKs based on DNNs are gaining momentum with the trend toward wider adoption of deep learning technologies on embedded systems. Developers can purchase/subscribe to the SDK and build their applications using the Application Programming Interface (API) that are made public. Legitimate users purchase access and then authenticate to use the resource services provided by the model owner. Since local SDKs are susceptible to unauthorized copying and distribution, they need to be controlled and traced to ensure that they are only accessible under legal authorization and can be traced and verified after being stolen.

Currently, various DNN model copyright protection schemes [4] prove an owner's copyright after the model is stolen, which is passive protection. To date, few studies have considered the active protection of DNN models through authorization control. Xue *et al.* [14] proposed a DNN copyright protection method using additional categories and image steganography techniques outside the dataset. Their method first selects a small number of images outside the original training dataset as watermark key samples. Then, the user's fingerprint is hidden in each watermark key sample through the least significant bit (LSB) technique: each user is assigned a unique fingerprint image so that the user's identity can be later verified. For legitimate users to use the DNN model, two conditions need to be met: 1) the DNN model classifies the watermark key samples into additional categories, and 2) the legitimate users' fingerprints are extracted from the watermark key samples. Although their method realized the authorization control of the DNN model, it has some disadvantages. First, a user must use the DNN model itself for authentication, which results in insufficient authorization control. Strict authorization control means that users must pass an identity authentication before they can access the DNN model. Second, the information embedded in the watermark key sample is only the user's fingerprint information, which does not establish a strong connection between the user and model owner. Additionally, the embedded information is not encrypted, which means it is insecure, and malicious attackers may forge a legitimate user identity to access the DNN model.

A DNN model active copyright protection and traceability (ACPT) framework is proposed. It is based on the work of Xue *et al.* [14] and the authorization control strategy and image perceptual hash algorithm; moreover, it uses the authorization control center constructed by the detector and verifier. In the framework, the detector detects whether the key image input by the user is legal, and the authenticator verifies whether the user identity information is legal. The ACPT framework implements strict authorization control, establishes a strong connection between users and model owners, and improves the framework security. Furthermore, the ACPT framework realizes the traceability of the stolen DNN model.

In short, the contributions of this study are as follows:

- A PCPT framework is proposed. The PCPT framework utilizes the black-box DNN watermarking technique using additional classes as trigger sets, which minimizes (or even eliminates) the effect of distortion of the original decision boundary. Moreover, the additional classes do not exist in the unlabeled model, and thus, the framework achieves a zero false-positive rate for the unlabeled model. The PCPT framework uses video framing technology to make it easier for the DNN model to learn the characteristics of the trigger set and simultaneously to reduce the protection difficulty of the trigger set. Together with blockchain technology, the security of the PCPT framework is improved. Additionally, to our knowledge, this is the first time an image perceptual hash algorithm has been used to uniquely associate trigger sets with owner identities.

- An ACPT framework is proposed. The ACPT framework utilizes the authorization control center constructed by the detector and validator to realize stricter authorization control, to establish a strong connection between a user and model owner, and to improve the framework security. Moreover, the ACPT framework realizes the traceability of the stolen DNN model.

- Experiments were conducted on the PCPT framework using the LeNet5, VGG16, GoogLeNet, ResNet18, and DenseNet121 models on the MNIST and CIFAR10 datasets to verify the effectiveness of the PCPT framework for DNN model tracking and traceability, as well as robustness, to model modification attacks. Simultaneously, the VGG16, GoogLeNet, and ResNet18 models were used as examples to conduct experiments on the ACPT framework, proving the effectiveness of the ACPT framework's authorization control and traceability performance. Furthermore, the hiding of key samples was better concealed in the ACPT framework.

The rest of the paper is structured as follows. Section II combs the relevant research work, and section III briefly introduces the background knowledge closely related to our work. Section IV discusses the threat model, Section V introduces the PCPT framework and its evaluation experiments, and Section VI introduces the ACPT framework and its evaluation experiments. Finally, Section VII presents the conclusions, and an outlook for future research is proposed.

## II. RELATED WORK

In this section, we sort out the passive and active copyright protection schemes of DNN model.

1) **Passive Protection Scheme**

The passive copyright protection scheme for DNN models was first proposed by Uchida *et al.* [5], and it used watermarking to handle the DNN model. According to different application scenarios, DNN watermarking can be divided into four categories: white-box, black-box, gray-box, and null-box watermarks. Herein, black-box DNN watermarking is mainly analyzed. In black-box DNN watermarking, the model owner constructs a trigger set through specific inputs and outputs to transform the model. During watermark verification, the model owner uses the trigger set to verify ownership. The existing black-box DNN watermarking work can be further divided according to the different construction methods of the trigger set.

a) The trigger set is constructed with only label changes. Adi *et al.* [7] constructed a trigger set using a set of abstract images and labels that are inconsistent with the image content, and they randomly assigned labels to trigger set images. After the trigger set is input, the watermarked DNN model outputs a specific label to verify the copyright of the model.

b) A trigger set is constructed using embedding information and label changes in the original samples. Zhang *et al.* [8] studied three watermark generation algorithms, including embedding meaningful text content and meaningless noise into image samples as watermarks and assigning incorrect labels to irrelevant samples as watermarks. Based on their work, a backdoor watermarking method is proposed to embed watermarks into target models. Guo *et al.* [13] proposed a watermarking framework that fine-tunes the initial model using both the original and modified datasets so that the trained watermark-containing model will run in a predefined special pattern when it encounters any input embedded with the copyright owner's signature.

c) The trigger set is constructed by adding a new class. Zhong *et al.* [15] watermarked models by adding new class labels to carefully crafted key samples during training. By modifying the task of predicting (N−1) different classes from the original target model to predict N different classes, a watermark-free model cannot output a nonexistent class label. The above scheme for DNN model copyright protection provides a reference for the PCPT framework. However, in the above scheme, the watermark is the same in all copies of the model. Thus, after the DNN model is leaked, the owner cannot trace the source.

2) **Active Protection Scheme**

The active protection scheme for the DNN model is addressed using authorization control. Chen *et al.* [16] used an additional anti-piracy conversion module to verify the legitimacy of users, providing authorization controls for trained DNNs so that only authorized users can use them correctly. Fan *et al.* [17] exploited passports to control the performance of DNN models, whose performance either remained unchanged in the presence of valid passports or significantly deteriorated due to modified or counterfeit passports. Chakraborty *et al.* [18] implemented authorization control for DNN models using a hardware-assisted approach, which relied on a trusted hardware device (as a root of trust) to store each user's key, ensuring that only trusted hardware devices (with keys embedded on the chip) run the intended deep learning application using the published model. However, their method is expensive for commercial applications. Furthermore, the above active authorization control methods do not support user authentication management, making them unsuitable for commercial applications. Additionally, through the active protection of a DNN model, all users' copies of the model are identical. Thus, once a DNN model is leaked, the owner cannot trace the source.

In summary, few studies have been conducted to date on the traceability of DNN models after they have been stolen. The KeyNet framework proposed by Jebreel *et al.* [12] addresses the attribution problem of DNN models, but the framework yields a small number of false-positives when used in models without watermarks. Therefore, the PCPT and ACPT frameworks are proposed to address the problem of traceability after a DNN model is stolen.

## III. BACKGROUND

### A. DCT-PHA

Kalker *et al.* [19] first proposed "perceptual hashing" in 2001. The perceptual hash algorithm (PHA) maps multimedia data into a digest, which is one-way. Image perceptual hashing algorithms are used to generate "fingerprint" strings for images, mainly including the perceptual hash function based on the discrete cosine transform (DCT-PHA), the perceptual hash function based on the Marr–Hildreth operator, and the radial variance and block mean-based perceptual hash functions. In comparison, DCT-PHA produces better image recognition and resolution capabilities than other image perceptual hashing algorithms [20].

The processing of DCT-PHA can be divided into seven steps: 1) To facilitate DCT calculation, the image size is adjusted to 32 × 32. 2) The adjusted image is converted to grayscale. 3) DCT calculation is performed on the grayscale image. 4) The 8 × 8 low frequency region is maintained in the upper left corner of the image. 5) The average, *ave,* of all pixels in the low-frequency region is calculated. 6) By comparing the pixel value of each pixel in the low-frequency region with the size of *ave* via (1), an 8×8 binary matrix is obtained. 7) Following the order of the set, the outputs are combined to obtain the hash value of the image.

$$\begin{cases} 1, if\ pixel\ value > ave\ , \\ 0, \quad\quad other. \end{cases} \quad (1)$$

To strongly link a DNN model with an owner, researchers have been embedding the model owner information inside the trigger set image. This method has limited capacity to embed owner information and affects the quality of the original image. Therefore, the PCPT framework uses the idea of zero watermarks and adopts an image perceptual hash algorithm to uniquely associate an owner with a DNN model. This approach does not degrade the trigger set image quality, and detecting whether the trigger set image is protected is difficult for an attacker.

*B. Video framing*

Video framing is to divide the video into frames to extract and display. Why use video framing technology? First, after a digital video is framed, the content between adjacent frames is highly correlated. Therefore, compared to the general trigger set construction method, a very obvious feature of the trigger set constructed by video framing is that it has a strong correlation. This supports the DNN model to fully learn the trigger set characteristics. Second, the PCPT framework needs to allocate different trigger sets for users, increasing the difficulty of trigger set protection. If the trigger set is constructed using video framing technology, the model owner only needs to save a piece of video material that is not related to the DNN model as a private key.

*C. Reversible watermarking technology based on difference expansion*

Reversible watermarking algorithms can be divided into three categories: lossless compression based, histogram shift (HS) based and difference expansion (DE) based. The reversible watermarking algorithm based on lossless compression is one of the earliest reversible watermarking technologies. The algorithm uses lossless compression to obtain usable redundant space to embed watermark information in the visually insensitive area. The reversible watermarking algorithm based on histogram translation comes from the technology of histogram translation. This technology develops a histogram movement strategy for reversible data hiding by modifying the difference histogram of the image in the spatial domain.

Although the algorithm based on histogram translation has a good effect of hiding information, it is also difficult to embed a large amount of hidden information. To overcome this shortcoming, researchers propose reversible watermarking algorithms based on difference expansion, which are mainly divided into two categories: reversible watermarking algorithms based on pixel difference expansion and prediction error expansion. Tian first proposed a reversible watermarking algorithm based on difference expansion [21], embedding the watermark information into the extended difference. This method utilizes the characteristics of high correlation and small difference between adjacent pixels of the image to embed secret information. The algorithm has high embedding capacity and small distortion, but it does not take a good way to avoid pixel overflow, which affects the embedding capacity. Lee et al. proposed a reversible watermarking algorithm based on prediction error expansion (PEE) [22], which further improved the embedding capacity. Fig. 1 shows the comparison results of the capacity and imperceptibility of HS, DE and PEE reversible watermarking algorithms. The test image is Lena diagram. Compared with HS, PEE has greater embedded capacity at higher PSNR. Compared with DE, PEE is superior to DE in both capacity and imperceptibility. Therefore, the user key design of ACPT framework adopts the PEE algorithm, which not only improves the capacity of embedded user information, but also has good imperceptibility, and realizes lossless embedding and extraction.

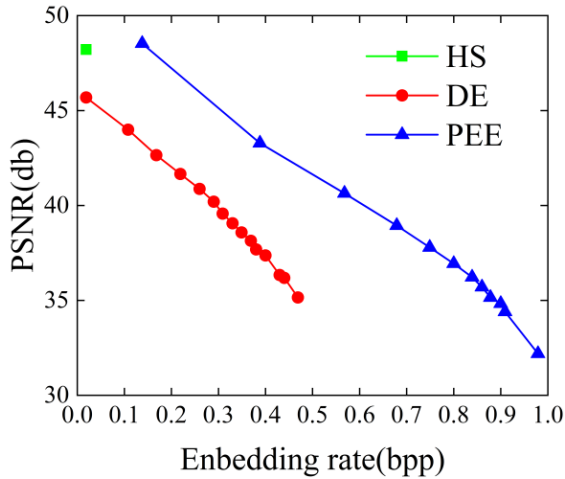

**Fig. 1.** Comparison results of PEE with HS and DE reversible schemes

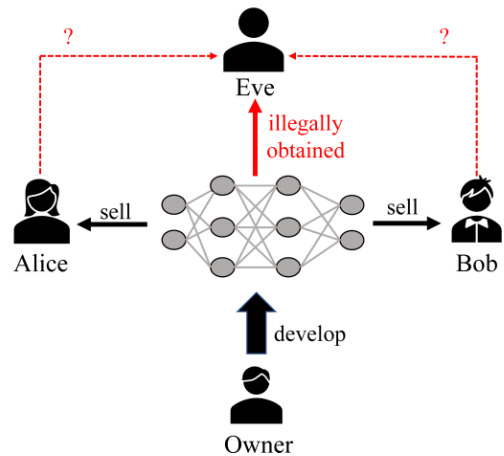

**Fig. 2.** Threat model

## IV. THREAT MODEL

As shown in Fig. 2, *Owner* is the model owner. *Owner* developed a DNN model and sold it to two (or more; two users are considered herein) users: *Alice* and *Bob*. *Alice* and *Bob* deploy the model for use by the specified population. Note that *Alice* and *Bob* are authorized users of *Owner* and only have the right to use the DNN model and do not have ownership. Thus, they need to bear the responsibility for copyright maintenance of the DNN model. After determining the function of the DNN model, the criminal *Eve* wants to steal a copy of the DNN model from *Alice* or *Bob* to distribute or provide services for profit. After *Eve* steals the DNN model, the main problems addressed herein are how the *Owner* traces the source of *Eve* stealing the DNN model and how the *Owner* pursues culpability of the authorized user for the leakage of intellectual property rights.

This study evaluates the copyright protection and traceability framework of DNN models. When the DNN model is stolen, *Owner* can lock the leak source of the model using the PCPT and ACPT frameworks, thus providing help for later accountability and rights protection.

## V. PCPT FRAMEWORK

In this section, the PCPT framework is proposed. The PCPT framework traces the origin of suspicious DNN models by validating watermarks in remote DNN services. To embed the watermark into a DNN model, the framework builds different trigger sets and corresponding predefined labels for different users; it converts trigger sets with predefined labels $W = \{w(i), Alice/Bob\}_{i=1}^{L}$ to an additional class and 10% of the original training data $D'_{train} = \{x(i), y(i)\}_{i=1}^{N}$ to fine-tune the training original DNN model $F$. Then, it generates the watermarked DNN models $F_{Alice}$ and $F_{Bob}$, as shown in (2). DNN models automatically learn and memorize patterns of embedded watermarks and predefined labels, and only DNN models protected by watermarks can generate predefined predictions. In the verification stage, after different trigger sets are input to a watermarked DNN model, a predefined additional class is output to achieve traceability.

$$F_{Alice}, F_{Bob} \leftarrow Train(F(D'_{train} \cup W)) \quad (2)$$

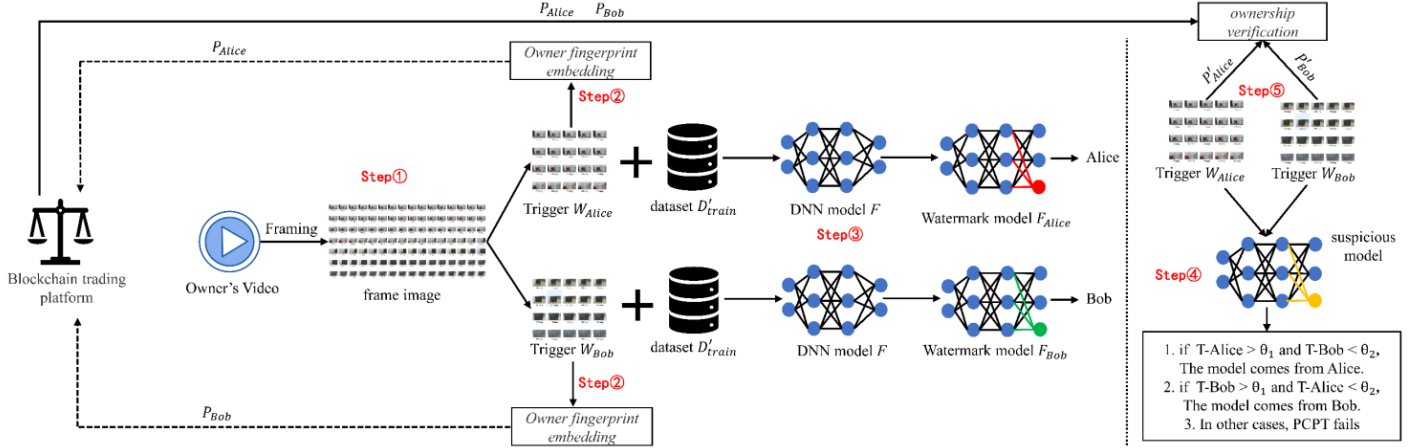

**Fig. 3.** PCPT framework

As shown in Fig. 3, the PCPT framework is mainly divided into five stages: trigger set generation, *Owner* fingerprint embedding, DNN watermark embedding, traceability verification, and ownership verification.

### A. Trigger set generation

The video framing technique is adopted to construct the trigger set. First, *Owner* shoots and produces a video that is rich in content and wherein different frame images can be differentiated. Then, the video is divided into frames, and some frame images with different contents are selected to construct the trigger sets $W_{Alice} = \{w(i), Alice\}_{i=1}^{L}$ and $W_{Bob} = \{w(i), Bob\}_{i=1}^{L}$. Herein, the number of users is two. If the number of users increases, the duration of the video material and the richness of the content should be simultaneously increased. The video footage is ultimately saved by the *Owner* being used as a key.

### B. Owner fingerprint embedding

To uniquely associate the trigger samples with the *Owner*, the *Owner* selects $L$ trigger images with high robustness against model modification attacks from the trigger set to embed their fingerprint. Fig. 4 shows the *Owner* fingerprint embedding process. First, the trigger image and *Owner* fingerprint image are processed using the DCT-PHA algorithm, and the hash values $P_1$ and $P_2$ of the image are obtained. Then, $P_1$ and $P_2$ are XORed to obtain $P$. Finally, storing $P$ in blockchain transactions [23] enables data protection [24].

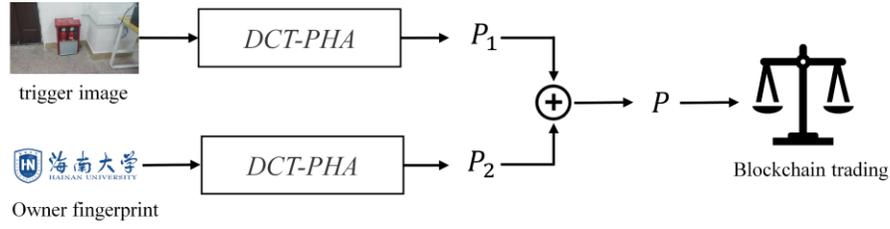

**Fig. 4.** *Owner* fingerprint embedding

*C. DNN watermark embedding*

To embed the watermark in the DNN model, the *Owner* uses different trigger sets $W_{Alice} = \{w(i), Alice\}_{i=1}^{L}$ and $W_{Bob} = \{w(i), Bob\}_{i=1}^{L}$ with 10% of the original training data $D'_{train} = \{x(i), y(i)\}_{i=1}^{N}$ to form a new training set $D_{Alice}$ and $D_{Bob}$ to fine-tune the DNN model $F$ and obtain the watermarked DNN models $F_{Alice}$ and $F_{Bob}$. The network structure of $F_{Alice}$ and $F_{Bob}$ is identical to that of F, except that the output layer adds a class (*Alice* or *Bob*) to the original one. The embedding and traceability algorithm of the DNN watermark is shown in Algorithm 1.

*D. Traceability verification*

When a suspicious DNN service is found, the *Owner* uses the trigger sets $W_{Alice} = \{w(i), Alice\}_{i=1}^{L}$ and $W_{Bob} = \{w(i), Bob\}_{i=1}^{L}$ to verify whether a watermark is present in the suspicious model. The test accuracies *T-Alice* and *T-Bob* of the trigger sets $W_{Alice} = \{w(i), Alice\}_{i=1}^{L}$ and $W_{Bob} = \{w(i), Bob\}_{i=1}^{L}$ are compared to assess traceability. Here, *Owner* sets three thresholds $\theta_1$, $\theta_2$ and $\theta_{Threshold}$. If *T-Alice* $> \theta_1$ and *T-Bob* $< \theta_2$, and $|T\text{-}Alice - T\text{-}Bob| \geq \theta_{Threshold}$, then *Alice* has leaked the model. Inversely, *Bob* has leaked the model. Beyond that, the PCPT framework fails.

**Algorithm 1 DNN Watermark Embedding and Traceability**

**Input:** Original DNN model $F$, original dataset $D_{train} = \{x(i), y(i)\}_{i=1}^{N}$, trigger set $W_{Alice} = \{w(i), Alice\}_{i=1}^{L}$, trigger set $W_{Bob} = \{w(i), Bob\}_{i=1}^{L}$, suspicious DNN model $F_{Sus}$, epoch, crossentropyloss, threshold $\theta_1$, $\theta_2$ and $\theta_{Threshold}$.

**Output:** Watermarked DNN model $F_{Alice}$ and $F_{Bob}$, Source of suspicious DNN model $F_{Sus}$ S.

1: $D'_{train} \leftarrow$ randomly choose 10% from $D_{train} = \{x(i), y(i)\}_{i=1}^{N}$
2: $D_{Alice} \leftarrow D'_{train} \cup W_{Alice}$
3: $D_{Bob} \leftarrow D'_{train} \cup W_{Bob}$
4: **for** i = 1, 2, 3 ..., epoch **do**
    $F_{Alice} \leftarrow$ Train ($F$ ($D_{Alice}$))
     crossentropyloss($F_{Alice}(W_{Alice})$, *Alice*)
   **end for**
5: **for** i = 1, 2, 3 ..., epoch **do**
    $F_{Bob} \leftarrow$ Train ($F$ ($D_{Bob}$))
     crossentropyloss($F_{Bob}(W_{Bob})$, *Bob*)
   **end for**
6: *T-Alice* = $F_{Alice}(W_{Alice})$ and *T-Bob* = $F_{Bob}(W_{Bob})$
7: **if** *T-Alice* $> \theta_1$ and *T-Bob* $< \theta_2$, and $|T\text{-}Alice - T\text{-}Bob| \geq \theta_{Threshold}$
    $S \leftarrow Alice$
   **if** *T-Bob* $> \theta_1$ and *T-Alice* $< \theta_2$, and $|T\text{-}Alice - T\text{-}Bob| \geq \theta_{Threshold}$
    $S \leftarrow Bob$
   **else**
    traceability failure
8: **return** $F_{Alice}$, $F_{Bob}$, S

*E. Ownership verification*

An advantage of utilizing the image perceptual hash algorithm is that criminals do not know whether the *Owner*'s data are protected and the image quality is not compromised. When the ownership of the DNN model needs to be proven, *Owner* first uses the trigger image to trigger the watermark in the DNN model to verify the copyright and then obtains $P'$ according to the method in 4.2. Finally, the smart contract is called to query the data $P$ stored in the blockchain transaction. If $P' = P$, the ownership verification is successful.

*F. Experiment analysis*

The performance of the PCPT framework is experimentally evaluated. All experiments are performed on the Google Colab platform. The graphics card is an NVIDIA Tesla P100, and the deep learning framework is PyTorch. First, the dataset and DNN model used in the PCPT experiments are introduced. Then, the effectiveness of the PCPT framework for tracing the source of the DNN model after being stolen is verified, and the impact of the PCPT framework embedded in the watermark on the accuracy of the original model is tested. Finally, the robustness of the PCPT framework to model modification attacks and the security and uniqueness of the PCPT framework are demonstrated.

1) **Experimental setup**

**Datasets and models.** The PCPT framework is evaluated on two datasets (MNIST and CIFAR10). Among them, the MNIST dataset adopts the LeNet5 network architecture, and the CIFAR10 dataset adopts the VGG16, GoogLeNet, ResNet18 and DenseNet121 network architectures. MNIST is used to train the LeNet5 model. The trigger set is constructed using 10% MNIST data, and the video frame segmentation image is used to fine-tune the trained LeNet5 model to embed the watermark. One hundred video frame images are assigned to each user to construct a trigger set. Similarly, the same settings are used for the CIFAR10 dataset and the VGG16, GoogLeNet, ResNet18 and DenseNet121 models.

**Parameter setup.** The thresholds are set to $\theta_1 = 85\%$, $\theta_2 = 60\%$ and $\theta_{Threshold} = 40\%$. In the fine-tuning embedding watermark stage, the cross-entropy loss is selected as the loss function to fine-tune the four models for 50 epochs.

2) **Validity**

The validity is evaluated to quantify whether *Owner* can trace the DNN model stolen by the criminals to *Alice* or *Bob*. As shown in TABLE I, when the watermarked DNN models $F_{Alice}$-LeNet5, $F_{Alice}$-VGG16, $F_{Alice}$-GoogLeNet, $F_{Alice}$-ResNet18, and $F_{Alice}$-DenseNet121 are assigned to *Alice*, the conditions of $T\text{-}Alice > \theta_1$ and $T\text{-}Bob < \theta_2$, as well as $|T\text{-}Alice - T\text{-}Bob| \geq \theta_{Threshold}$, are satisfied. When the watermarked DNN models $F_{Bob}$-LeNet5, $F_{Bob}$-VGG16, $F_{Bob}$-GoogLeNet, $F_{Bob}$-ResNet18, and $F_{Bob}$-DenseNet121 are assigned to *Bob*, the conditions of $T\text{-}Bob > \theta_1$ and $T\text{-}Alice < \theta_2$, as well as $|T\text{-}Alice - T\text{-}Bob| \geq \theta_{Threshold}$, are satisfied. As shown in Figure 5. Therefore, the PCPT framework effectively traces the source of the DNN models after they are stolen.

TABLE I. WATERMARK TEST ACCURACY OF THE WATERMARK MODEL

| DataSet | Model | WM accuracy | |
|---|---|---|---|
| | | *T-Alice* | *T-Bob* |
| MNIST | $F_{Alice}$-LeNet5 | 100% | 0 |
| | $F_{Bob}$-LeNet5 | 0 | 100% |
| CIFAR10 | $F_{Alice}$-VGG16 | 99% | 48% |
| | $F_{Bob}$-VGG16 | 11% | 100% |
| CIFAR10 | $F_{Alice}$ - GoogLeNet | 98% | 3% |
| | $F_{Bob}$ - GoogLeNet | 0 | 100% |
| CIFAR10 | $F_{Alice}$ - ResNet18 | 100% | 4% |
| | $F_{Bob}$ - ResNet18 | 2% | 100% |
| CIFAR10 | $F_{Alice}$- DenseNet121 | 100% | 34% |
| | $F_{Bob}$- DenseNet121 | 6% | 95% |

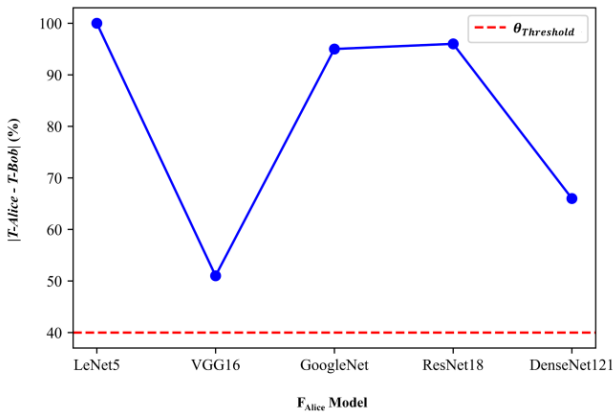 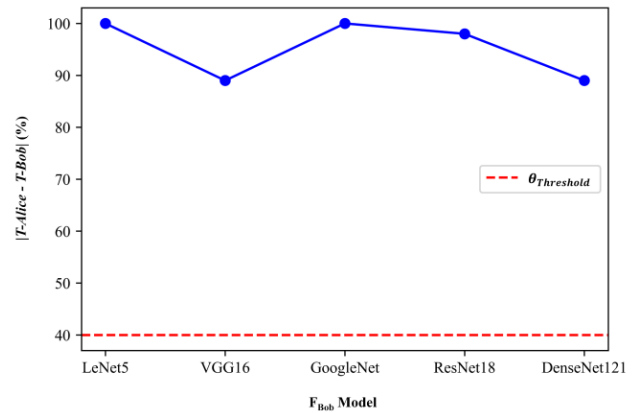

(a) (b)

**Fig. 5.** Absolute value of the difference between *T-Alice* and *T-Bob*

3) **Fidelity**

Fidelity requires our PCPT framework to watermark the original model without significant side effects on the main task of the original model. Ideally, the watermarked DNN model should be as accurate as the original DNN model. Fig. 6 shows the comparison between the watermarked models $F_{Alice}$ and $F_{Bob}$ and the original model $F$ in the test accuracy of the original task. The results show that the accuracy of the DNN model applying the PCPT framework decreases by 0.60% on average and is controlled to within 1% overall [13]. In the best case, We achieved an accuracy increase of 0.50%. Compared to Guo et al., who achieved a 0.23% decrease [13], the PCPT framework won by 0.73%. Thus, the side effects caused by the PCPT framework are well within the acceptable performance variation of the model and have no meaningful impact on the main task.

4) **Integrity**

Integrity requires that after the PCPT framework watermarks the original model, the generated trigger set will not generate false positives for the watermark-free model and can clearly distinguish between the watermarked model and the model without watermark. We used VGG16 and ResNet18 models to compare with the KeyNet framework proposed by Jebreel et al. [12]. The experimental results are shown in TABLE II. Among them, the model without watermark uses the same topology as the model with watermark. It can be seen that PCPT realizes zero false positive rate for the watermark free model because it adopts the idea of additional classes.

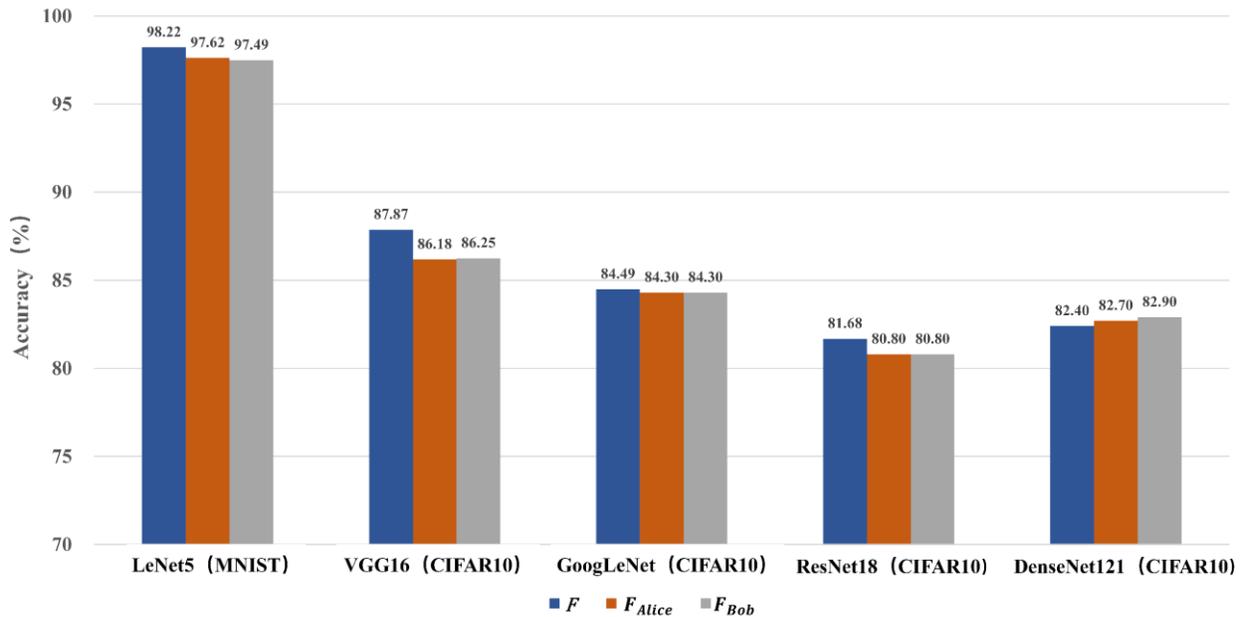

**Fig. 6.** Original task test accuracy for the original model $F$ and watermarking models $F_{Alice}$ and $F_{Bob}$

TABLE II. WATERMARK TEST ACCURACY IN THE MODEL WITHOUT WATERMARK

| DataSet | Model | WM accuracy | |
|---|---|---|---|
| | | KeyNet | PCPT |
| CIFAR10 | VGG16 | 7.92% | 0 |
| | ResNet18 | 18.92% | 0 |

5) **Robustness**

The resistance of the watermarked model to model modification attacks (fine-tuning and pruning attacks) is tested to measure the robustness of the watermarking method.

a) Fine-tuning attack. In this experiment, we split the test set of each dataset into two halves in the manner of Zhang et al. [8]. The first 50% is used to fine-tune the DNN model with watermarking, and the second 50% is used to evaluate the fine-tuned model. Then, the watermarked DNN model is fine-tuned and trained for 200 epochs. The experimental results are shown in TABLE III. For the MNIST dataset (LeNet-5), fine-tuning does not degrade the accuracy of watermarking and achieves the same effect as Zhang et al. [8]. For the CIFAR10 dataset, the greatest decrease in watermarking accuracy is in $F_{Alice}$-DenseNet121, which decreases by 11% after fine-tuning 50 epochs but still has a relatively high accuracy (89%), which is better than the 69.13% of Zhang et al. [8]. The set threshold is satisfied, so the PCPT framework is robust to fine-tuning attacks.

TABLE III. THE ORIGINAL TASK TEST ACCURACY (*T-ORIGINAL*) AND THE TEST ACCURACY ON DIFFERENT TRIGGER SETS (*T-ALICE, T-BOB*) OF THE WATERMARKING MODEL BEFORE AND AFTER THE FINE-TUNING ATTACK

| Number of Epochs | $F_{Alice}$-LeNet5 | | | $F_{Bob}$-LeNet5 | | |
|---|---|---|---|---|---|---|
| | T-Original | T-Alice | T-Bob | T-Original | T-Alice | T-Bob |
| 50 | 98.80% | 100% | 12% | 98.66% | 0 | 100% |
| 100 | 98.82% | 100% | 12% | 98.60% | 0 | 100% |
| 150 | 98.82% | 100% | 12% | 98.60% | 0 | 100% |
| 200 | 98.82% | 100% | 12% | 98.60% | 0 | 100% |

| Number of Epochs | $F_{Alice}$-VGG16 | | | $F_{Bob}$-VGG16 | | |
|---|---|---|---|---|---|---|
| | T-Original | T-Alice | T-Bob | T-Original | T-Alice | T-Bob |
| 50 | 85.73% | 96% | 1% | 86.14% | 20% | 100% |
| 100 | 85.39% | 97% | 26% | 86.08% | 17% | 99% |
| 150 | 86.02% | 96% | 1% | 85.98% | 14% | 100% |
| 200 | 85.90% | 97% | 16% | 85.98% | 11% | 100% |

| Number of Epochs | $F_{Alice}$-GoogLeNet | | | $F_{Bob}$-GoogLeNet | | |
|---|---|---|---|---|---|---|
| | T-Original | T-Alice | T-Bob | T-Original | T-Alice | T-Bob |
| 50 | 84.33% | 98% | 5% | 84.17% | 0 | 98% |
| 100 | 84.19% | 97% | 2% | 84.25% | 0 | 96% |
| 150 | 84.23% | 98% | 5% | 84.23% | 7% | 98% |
| 200 | 84.49% | 98% | 6% | 84.41% | 0 | 97% |

| Number of Epochs | $F_{Alice}$-ResNet18 | | | $F_{Bob}$-ResNet18 | | |
|---|---|---|---|---|---|---|
| | T-Original | T-Alice | T-Bob | T-Original | T-Alice | T-Bob |
| 50 | 80.78% | 98% | 0 | 81.21% | 0 | 98% |
| 100 | 80.99% | 98% | 0 | 80.70% | 0 | 97% |
| 150 | 80.85% | 94% | 0 | 80.91% | 0 | 97% |
| 200 | 81.03% | 98% | 0 | 80.98% | 0 | 98% |

| Number of Epochs | $F_{Alice}$-DenseNet121 | | | $F_{Bob}$-DenseNet121 | | |
|---|---|---|---|---|---|---|
| | T-Original | T-Alice | T-Bob | T-Original | T-Alice | T-Bob |
| 50 | 83.44% | 89% | 16% | 83.38% | 90% | 8% |
| 100 | 83.22% | 89% | 15% | 83.36% | 93% | 9% |
| 150 | 83.32% | 95% | 31% | 83.44% | 91% | 9% |
| 200 | 83.44% | 95% | 33% | 83.54% | 90% | 7% |

b) Pruning attack. In this experiment, the global pruning method is used to prune the weight parameters of the watermarking model. Fig. 7 shows the effect of model compression on the original task and trigger set test accuracies under different pruning rates. As shown in the figure, when pruning approximately 80% of the parameters of the LeNet5, ResNet18 and DenseNet121 models and pruning approximately 50% of the parameters of the VGG16 and GoogLeNet models, the conditions $T\text{-}Alice > \theta_1$ and $T\text{-}Bob < \theta_2$, as well as $|T\text{-}Alice - T\text{-}Bob| \geq \theta_{Threshold}$ in Alice's Model, and $T\text{-}Alice < \theta_2$ and $T\text{-}Bob > \theta_1$, as well as $|T\text{-}Alice - T\text{-}Bob| \geq \theta_{Threshold}$ in Bob's Model, are still satisfied. Thus, the PCPT framework is valid. As the pruning rate increases, the test accuracy of the DNN model on the trigger set will decrease. However, the attacker will not delete more than 50% of the model parameters because when the deletion rate exceeds 50%, the test accuracy of the model will sharply decrease. In other words, the attacker cannot prune the embedded watermark while maintaining the normal performance of the model. Therefore, the PCPT framework is robust to pruning attacks.

6) **Security**

The security of the PCPT framework from the perspective of *Eve*'s illegal ownership claims on DNN models is evaluated.

***Eve* forges an illegal trigger set similar to the original trigger sample to make an ownership claim.** The premise of this case is that *Eve* discovers a hidden pattern of triggering samples in the DNN model. Twelve images are randomly selected from the illegal trigger set forged by *Eve* for different models, as shown in Fig. 8, and watermark verification is performed. The experimental results are shown in TABLE IV. The table shows that the trigger set forged by *Eve* is not always able to trigger the DNN watermark. Additionally, the structural illegal trigger set forged by *Eve* and the *Owner*'s 100 legal trigger set images. The mean SSIM of the comparison results is shown in TABLE IV. The mean value of SSIM is the highest at 0.5017, demonstrating that the trigger set forged by *Eve* is not the frame image in the *Owner* key (video material) (SSIM < 0.9). Therefore, even if *Eve*'s forged trigger set successfully triggers the watermark in the DNN model, *Owner* can still prove that *Eve*'s forged trigger set illegally claims ownership.

*Eve fakes a new trigger set to claim ownership.* In this case, *Eve* is unaware of the hidden trigger sample patterns in the DNN model. However, *Eve* slipped their watermark into the DNN model by faking a new trigger set. During judicial authentication, two watermarks are present in the DNN model, and the ownership of the DNN model is ambiguous. However, the fingerprint information was embedded in the original trigger set by the *Owner* and stored in the blockchain transaction. Because the storage time and content written in the blockchain transaction cannot be tampered with, it is used in ownership authentication. Thus, ambiguity is avoided, and strong validation is provided to the *Owner*. Even if *Eve* uses the same operation to store their fingerprint on the blockchain, *Owner* stored it earlier than *Eve*, which still confirms the fact that *Eve* is forging the trigger set to illegally claim ownership.

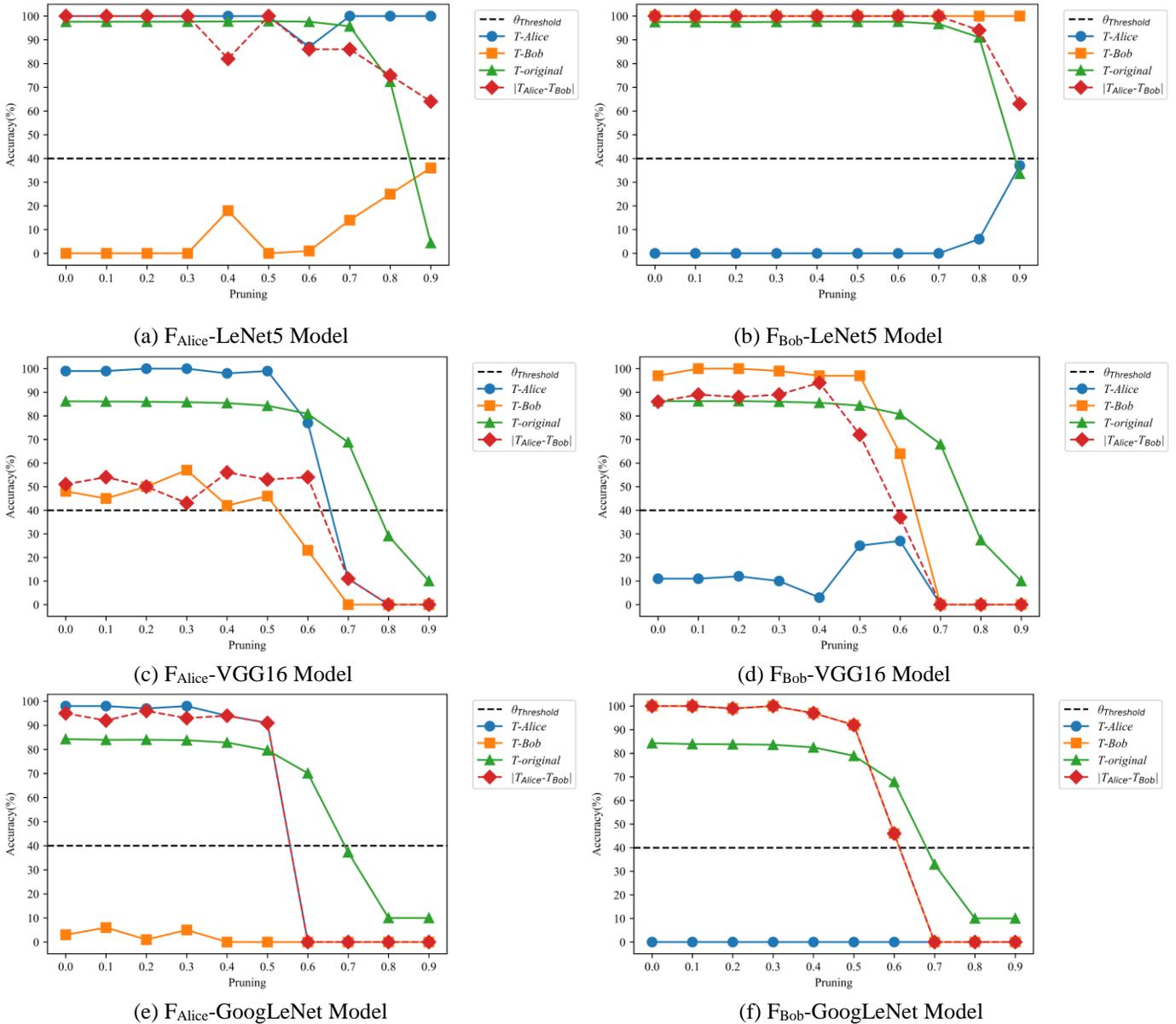

(a) $F_{Alice}$-LeNet5 Model  (b) $F_{Bob}$-LeNet5 Model

(c) $F_{Alice}$-VGG16 Model  (d) $F_{Bob}$-VGG16 Model

(e) $F_{Alice}$-GoogLeNet Model  (f) $F_{Bob}$-GoogLeNet Model

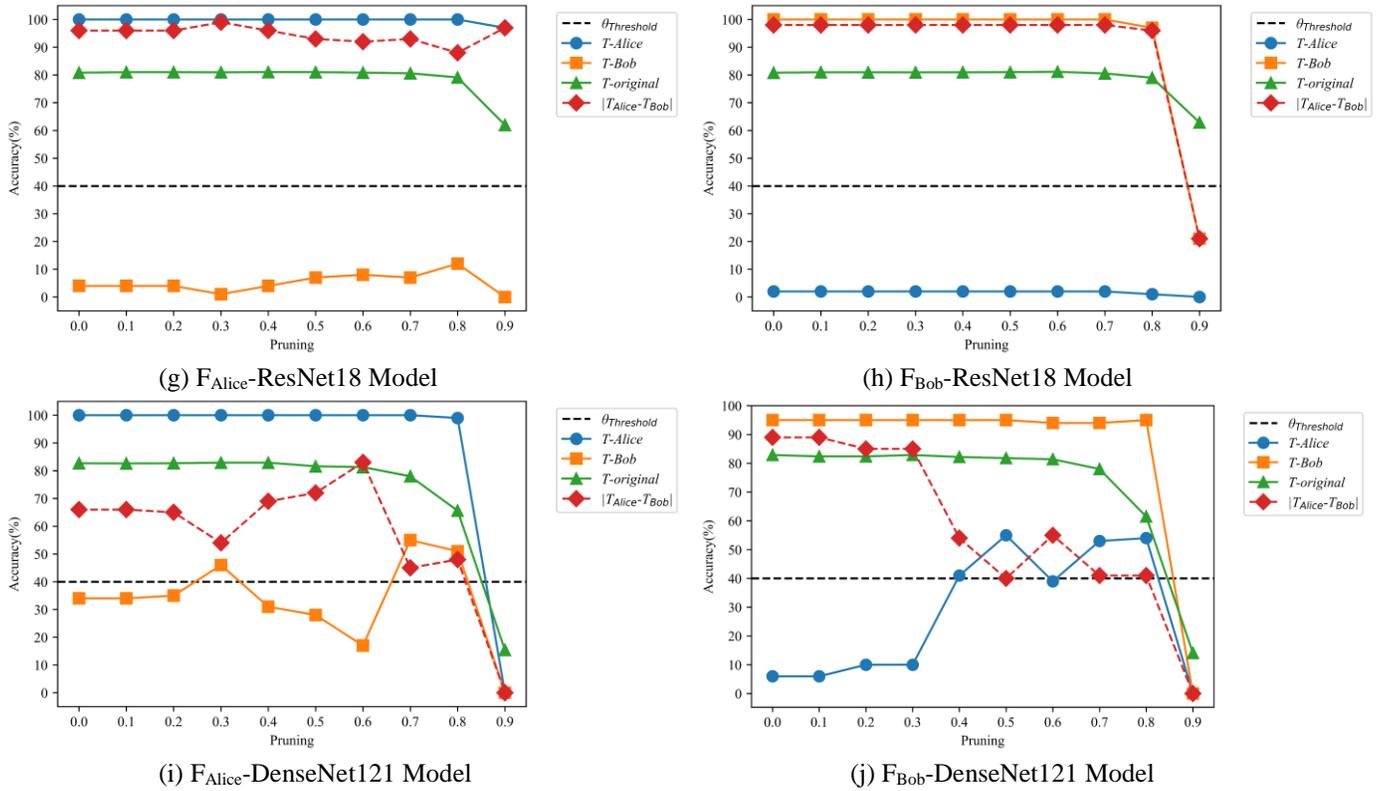

(g) $F_{Alice}$-ResNet18 Model  
(h) $F_{Bob}$-ResNet18 Model  
(i) $F_{Alice}$-DenseNet121 Model  
(j) $F_{Bob}$-DenseNet121 Model  

**Fig. 7.** The original task test accuracy (*T-Original*) of the watermarking model under different pruning rates and the test accuracy on different trigger sets (*T-Alice* and *T-Bob*)

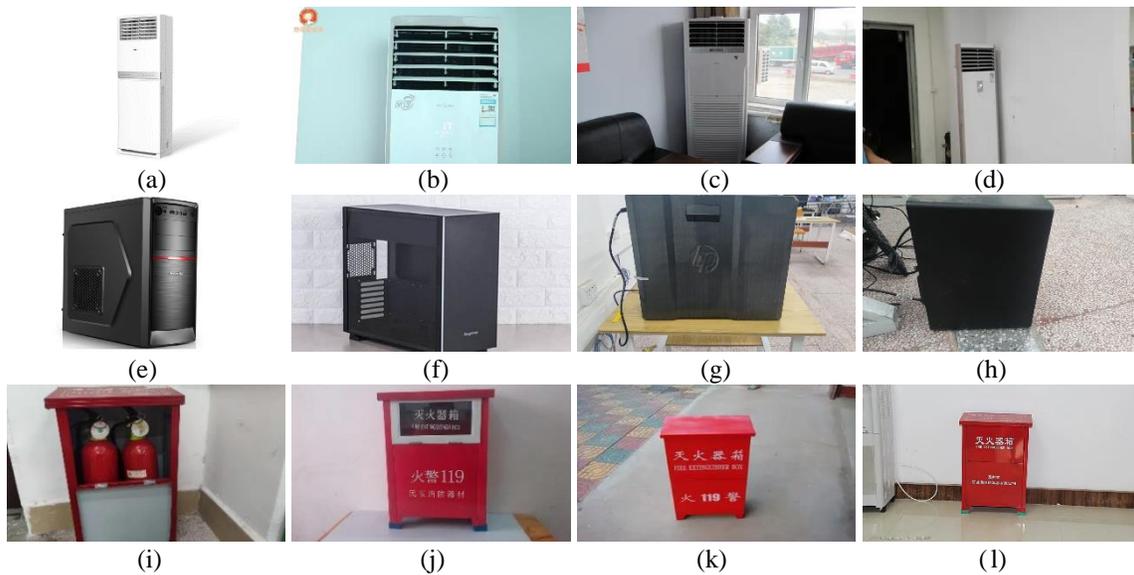

**Fig. 8.** The 12 illegal trigger sets. (a)–(d) are the trigger sets faked for *Alice*'s LeNet5 model, (e)–(h) are the trigger sets faked for *Bob*'s LeNet5, VGG16, GoogLeNet, ResNet18, and DenseNet121 models, (i)–(l) are the fake trigger sets for *Alice*'s VGG16, GoogLeNet, ResNet18, and DenseNet121 models.

TABLE IV. "$F_{Alice}$-LeNet5 watermark" INDICATES THE RESULT OF TRIGGERING THE DNN WATERMARK IN THE $F_{Alice}$-LeNet5 MODEL BY *EVE*'S FORGED TRIGGER SET ("√" INDICATES SUCCESSFUL TRIGGERING AND "×" INDICATES TRIGGER FAILURE), AND THE OTHERS ARE SIMILAR. "SSIM" DENOTES THE MEAN OF THE SSIM COMPARISON RESULTS BETWEEN EVE'S FORGED ILLEGAL TRIGGER SET AND THE OWNER'S 100 LEGAL TRIGGER SET IMAGES

| Fake trigger set | Fig. 8 (a) | Fig. 8 (b) | Fig. 8 (c) | Fig. 8 (d) |
|---|---|---|---|---|
| SSIM | 0.4528 | 0.3002 | 0.3491 | 0.3844 |
| $F_{Alice}$-LeNet5 watermark | √ | × | × | √ |
| Fake trigger set | Fig. 8 (e) | Fig. 8 (f) | Fig. 8 (g) | Fig. 8 (h) |
| SSIM | 0.4972 | 0.4644 | 0.5017 | 0.2746 |
| $F_{Bob}$-LeNet5 watermark | √ | √ | √ | √ |
| Fake trigger set | Fig. 8 (i) | Fig. 8 (j) | Fig. 8 (k) | Fig. 8 (l) |
| SSIM | 0.3434 | 0.3949 | 0.4245 | 0.4817 |
| $F_{Alice}$-VGG16 watermark | × | × | × | × |
| $F_{Alice}$-GoogLeNet watermark | × | √ | √ | √ |
| $F_{Alice}$-ResNet18 watermark | × | × | × | × |
| $F_{Alice}$-DenseNet121 watermark | × | × | × | × |
| Fake trigger set | Fig. 8 (e) | Fig. 8 (f) | Fig. 8 (g) | Fig. 8 (h) |
| SSIM | 0.4972 | 0.4644 | 0.5017 | 0.2746 |
| $F_{Bob}$-VGG16 watermark | × | × | √ | × |
| $F_{Bob}$-GoogLeNet watermark | × | √ | × | × |
| $F_{Bob}$-ResNet18 watermark | × | √ | × | √ |
| $F_{Bob}$-DenseNet121 watermark | × | × | × | × |

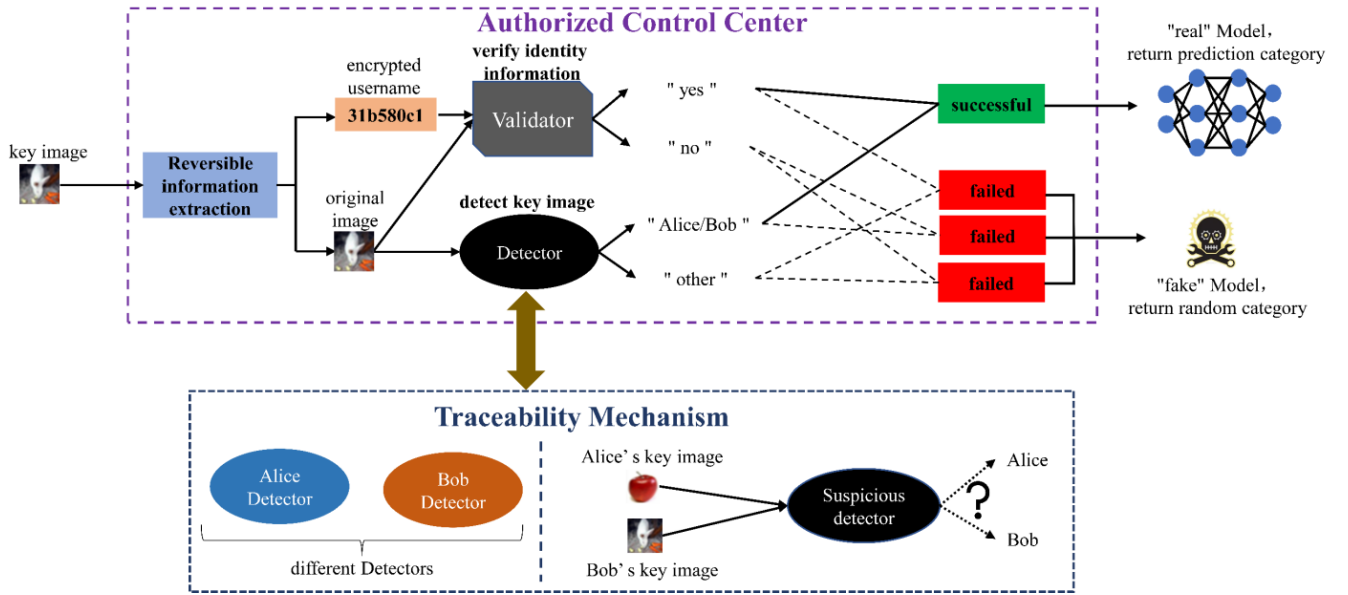

**Fig. 9.** ACPT framework

## VI. ACPT FRAMEWORK

As described in Section I, legitimate users who purchase access to a DNN model are authenticated to use the services provided by the model owner. Local SDK services are vulnerable to unauthorized replication and distribution. Therefore, this section discusses authorization controls for users to ensure that they are only accessible under legitimate authorization. Additionally, traceability verification can be performed after the models are stolen.

For *Owner* control of legitimate user access to DNN models, the DNN model ACPT framework is proposed. First, the ACPT framework builds an authorization control center using a detector and validator. Then, it packages the DNN model to be protected with the authorization control center ($DNN_{AC}$) and distributes it to users, thereby realizing active authorization control and user identity management of the DNN model. Second, the ACPT framework builds different authorization control centers $AC = \{AC_{Alice}, AC_{Bob}\}$ for different users to trace the source of the suspicious DNN models after the DNN model is stolen. It is worth noting that in the ACPT framework, Alice and Bob represent two groups of users. Each group of users has 100 subusers (corresponding to the 100 "apple" or "rabbit" images used to train the detector, each image with an encrypted username forming a subuser). In the verification stage, different keys are input into $DNN_{AC}$ to verify whether the DNN model can be correctly used to achieve traceability.

As shown in Fig. 9, the ACPT framework is divided into two parts: the authorization control center and traceability mechanism. The authorization control center comprises a detector and verifier, which realizes the active copyright protection of the DNN model, and the traceability mechanism realizes the traceability verification of the DNN model.

*A. User's key design*

The importance of information protection is increasing in today's society. The most common security scheme to control the authentication process for legitimate users to access information properly is passwords, which generally use a combination of numbers and letters. This authentication technique is called text-based authentication, and it has several drawbacks. For example, users habitually choose short and meaningful (e.g., date of birth, name, license plate number) to remember their passwords [25]. However, such passwords can be easily broken. If longer and meaningless passwords are set for users, it increases the security but also creates difficulties for users to remember them. Therefore, an additional mechanism based on images to improve password difficulty and memorability is a better solution. The combination of text and images can improve the resistance to certain password attacks, such as brute force and shoulder peek attacks [26].

We propose a hybrid authentication scheme that integrates text and image passwords. In addition to security aspects, the proposed scheme increases memorability because it does not require users to remember long and complex passwords. Thus, with the scheme, strong passwords can be created for users without sacrificing usability. The key, which is made of images for the hybrid authentication, is generated as follows.

1) An encrypted username for each legitimate user is generated. An example of the process of generating the username can be illustrated as follows. First, we assume that the real username is **user1** and the fingerprint information of Owner is **HN**. **user1** and **HN** are combined together as **HNuser1** and then encrypted with sha256 to obtain a 64-bit string m. The key $K_1$ is used to randomly choose 8 bits from m as an encrypted username, and it enhances the relation between Owner and the legitimate user. The probability that an illegal user forges a legitimate encrypted username without knowing Owner's fingerprint is $\frac{1}{2,821,109,907,456}$, so it can be assumed that it is impossible for an illegal user to successfully forge a legitimate encrypted username.

2) The encrypted username is embedded in the selected image with a reversible watermarking technology based on prediction error expansion, and the embedding and extraction process is shown in Fig. 10.

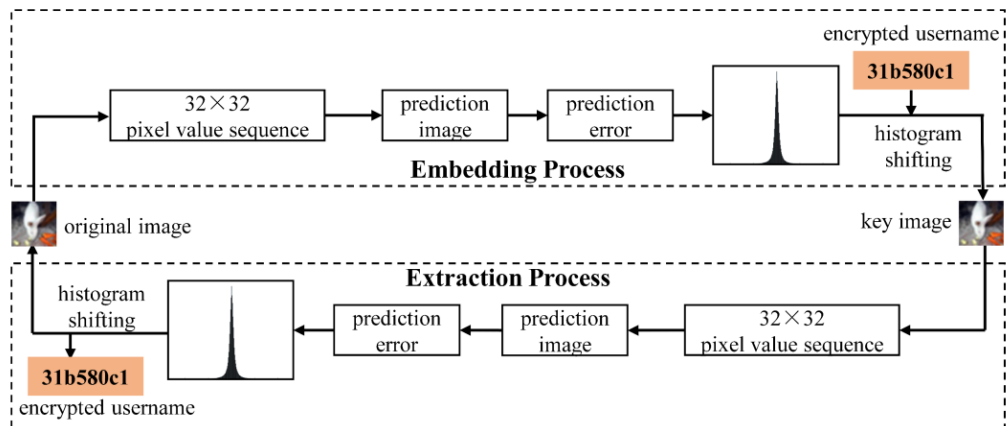

**Fig. 10.** Embedding and extraction process for reversible watermarking technology based on prediction error expansion

**Embedding process.** 1) The original image pixel value sequence is obtained, and the center pixel is predicted by the combination of four nearest neighbor pixels to obtain the predicted value of the original image, i.e., the predicted image. 2) The error between the predicted image sequence and the original image sequence is calculated to obtain the prediction error. 3) A prediction error histogram is generated from the prediction error, and then, the extended histogram translation method is used to embed the encrypted username into the image to obtain the key image.

**Extraction process.** 1) The sequence of key image pixel values is obtained, and the center pixel is predicted by a combination of four nearest neighbor pixels to obtain the predicted value of the key image. 2) The error between the sequence of predicted values and the sequence of key images is calculated to obtain the prediction error. 3) A prediction error histogram is generated from the prediction error, and then, the information is extracted from the image using extended histogram shifting. Finally, the image is recovered to obtain the encrypted username and the original image.

*B. Authorized control center*

The authorization control center comprises a detector and validator. After the user enters the image as a key, the encrypted username and the original image are first obtained by a reversible watermarking technology based on prediction error expansion extraction algorithm and then transferred to the detector and verifier for processing. The design methods of the detector and validator are introduced below.

1) **Detector**

The design idea of the detector stems from the attack method of Hitaj *et al.* [27]. In the ACPT framework, the detection of key images input by users is modeled as a binary classification problem, and the detector is trained to distinguish between legitimate and illegitimate users (Fig. 9). The detector takes the key images input by the user as the input and outputs two kinds of results: 1) a valid user key (*Alice* or *Bob*) and 2) an illegal user key (*other*).

2) **Validator**

The authenticator takes the encrypted username and the original image as input, and after processing, it generates the verification result, as shown in Algorithm 2. First, the encrypted username is converted into a 64-bit binary string $m_1$. Second, the 64-bit binary string $m_2$ is obtained after the original image is processed by DCT-PHA. Finally, the verification information $I$ is obtained by performing XOR on $m_1$ and $m_2$. If $I \in Q = \{I_1, I_2, I_3, ... , I_i\}$, the authentication is successful; otherwise, it fails. Here, $Q$ is the legitimate user identity information base.

The step of processing the original image with DCT-PHA in the validator is to enhance the sensitivity of the key image. If the encrypted username is leaked, the attacker embeds it in the illegal original image through the public reversible watermarking technology based on prediction error expansion algorithm for illegal access. In this case, the attacker's authentication still fails.

**Algorithm 2 Validator Process**

**Input:** *encrypted username, original image*.
**Output:** *"yes"* or *"no"*.
  1: $m_1 \leftarrow$ Binary(*encrypted username*)
  2: $m_2 \leftarrow$ DCT-PHA (*original image*)
  3: $I \leftarrow m_1 \oplus m_2$
  4: **if** $I \in Q = \{I_1, I_2, I_3, ... , I_i\}$
        **return** "yes"
    **else**
        **return** "no"

3) **Authorization control**

If the detector detects that a user's key is valid and the authenticator successfully verifies the user's identity information, the user passes the authorization control. A legitimate user will use the "true" model and obtain the predicted class of the DNN model. Otherwise, the user authentication fails, and the illegal user will use the "fake" model to obtain a random category of the DNN model. In this setting, the probability of an illegal user acquiring the correct predicted class is $1/N$, where $N$ is the number of labels present in the model. Note that the detector may employ a strategy of always rejecting queries from illegal users. However, the design of the random return category could make it difficult for the attacker (illegal user) to realize the existence of the authorized control center; thus, the return of the random category to the illegal user query is proposed.

*C. Traceability mechanism*

An authorized control center built by detectors and validators enables the active copyright protection of DNN models. Based on this, the ACPT framework builds different authorization control centers for different users and establishes a traceability mechanism to realize traceability after the DNN model is stolen. Specifically, the ACPT framework utilizes two different datasets (such as the

"apple" and "rabbit" datasets) as its key images to build two different detectors, which are assigned to users *Alice* and *Bob*. When the DNN model is leaked, the *Owner* uses different key images to authenticate the suspicious DNN and to examine the test set to determine the source of the suspicious DNN model.

*D. Experiment analysis*

The performance of the ACPT framework is experimentally evaluated. All experiments are performed on the Google Colab platform. The graphics card is an NVIDIA Tesla P100, and the deep learning framework is PyTorch. First, the detector setup in the ACPT framework is introduced, and the performance of the detector is evaluated. Then, the authorization control performance of the ACPT framework is verified, and the concealment of the key samples is examined. Finally, traceability validity verification is conducted.

1) **Detector setup and performance**

*Datasets and models.* The detector adopts the LeNet5 network architecture. Hundred images from the "apple" dataset and 100 other images (such as "cat", "chair" and "bee", etc.) from CIFAR100 are selected to train the $Detector_{Alice}$ assigned to user *Alice*, and 100 images from the "rabbit" dataset and 100 other images are selected to train the detector assigned to user $Bob\ Detector_{Bob}$.

*Parameter setup.* The detector is trained using an Adam optimizer for 50 epochs, and the cross-entropy loss is chosen as the loss function.

*Performance.* The detection accuracy of the trained detector on the key image is analyzed. The detection accuracy of the detector on the key image reaches 100%, which meets the requirements of legal user authorization control.

2) **Comparison with current work**

**Concealment of embedded information.** The concealment of the embedded information can be evaluated in both subjective and objective ways.

Subjective way. Fig. 11 shows an example of a key generated by reversible watermarking technology based on prediction error expansion. The first line is the original image and the second line is the key image. The embedded key has essentially no effect on the image quality, which shows good concealment.

Objective way. The average mean square error (MSE) is used to evaluate the concealment of embedding. The average MSE of the 100 keys generated by the four methods is shown in TABLE V. The results show that the average MSE of the reversible watermarking technology based on prediction error expansion we used is 0.701, which is close to the LSB method proposed by Xue et al. [14]. However, compared with other methods, the proposed method still has advantages and has little impact on the image quality. This indicates that the embedded information has concealment. In addition, although ACPT framework uses reversible watermark technology based on prediction error expansion to embed information in the key image, its concealment is slightly less than that of LSB method [14], the advantage of lossless recovery can make it better applied to the detector input of the authorization control center.

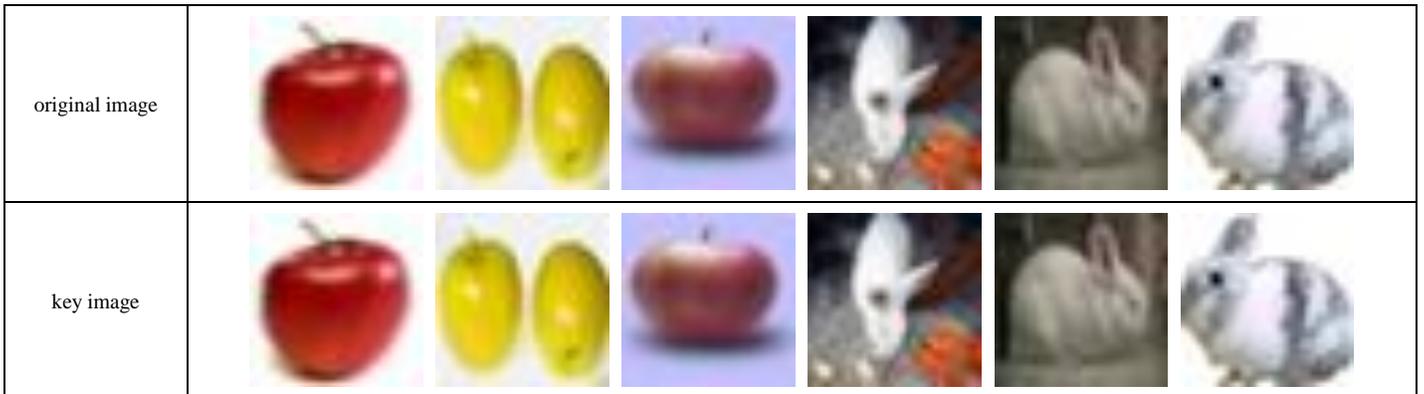

**Fig. 11.** Example of original image vs. key image

TABLE V. COMPARISON OF THE AVERAGE MSE OF DIFFERENT METHODS

| Dataset | Watermark type | Average MSE |
|---|---|---|
| CIFAR10 | Logo[8] | 203.58 |
| | Noise[8] | 298.96 |
| | LSB[14] | 0.015 |
| | Ours | **0.701** |

**Performance of authorization control.** Strict authorization control means that users must be authenticated before they can access the DNN model. The scheme of Xue et al. [14] is an example in which user authentication needs to be implemented with the help of the DNN model itself; this leads to less strict authorization control. The ACPT framework achieves strict authorization control by building an authorization control center that enables user authentication without the help of the DNN model itself.

**Security.** The ACPT framework builds on Xue et al.'s scheme [14] by embedding a string that uniquely represents the user's identity information as the user's fingerprint into the trigger set and further enhances security by encrypting the user's fingerprint embedded in the trigger set carrier. As described in Section VI.A, this makes it virtually impossible for an illegal user to obtain DNN model resources by forging a legitimate encrypted username.

TABLE VI. PERFORMANCE COMPARISON WITH THE SCHEME PROPOSED BY XUE ET AL. [14]

| Norm | Xue et al. [14] | ACPT |
|---|---|---|
| Strict authorization control | × | √ |
| User and model owner connections | × | √ |
| User fingerprint encryption | × | √ |
| Traceability | × | √ |

**Traceability.** Compared with the scheme of Xue et al. [14], ACPT also establishes a traceability mechanism by constructing different authorization control centers for different users to achieve traceability after the DNN model is stolen, as described in Section VI.C.

The above comparison is shown in TABLE VI.

3) **Authorization control performance**

Fig. 12 shows the test accuracy for the authorized and unauthorized use of the VGG16, GoogLeNet, and ResNet18 models on the CIFAR10 dataset. The results show that the test accuracy of authorized users on these three models is 87.87% (VGG16), 84.49% (GoogLeNet), and 81.68% (ResNet18). However, the test accuracies of the unauthorized users are 9.26% (VGG16), 10.50% (GoogLeNet), and 10.00% (ResNet18), which is an average 74.76% less than the performance of authorized users. Therefore, the proposed ACPT framework realizes active authorization control and effectively prevents the illegal use of the DNN model.

4) *Traceability verification*

In the experiment, the *Owner* distributed two copies of the DNN model for users *Alice* and *Bob*, each with its corresponding authorization control center $AC = \{AC_{Alice}, AC_{Bob}\}$. The VGG16 model is taken as an example, and it is assumed that *Bob* leaks the VGG16 model. Next, the traceability mechanism in the ACPT framework is employed to try to find the leaker.

- A key image is taken from both the "apple" and "rabbit" keys.

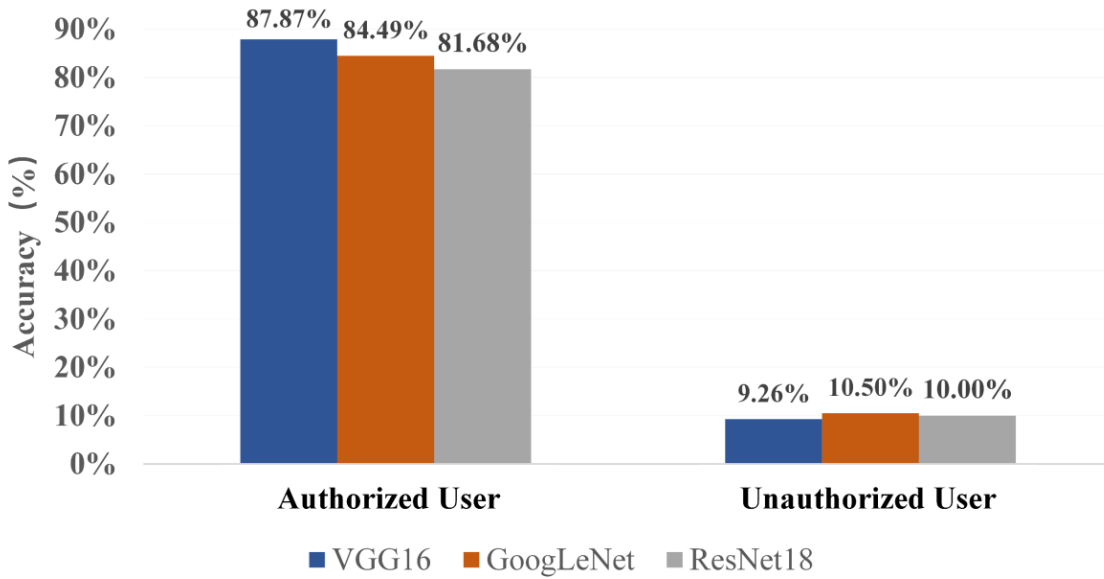

**Fig. 12.** Authorization control performance

- The suspicious model is authenticated using the two keys, and then, an image is selected from the 10 classes of CIFAR10 to test the suspicious model.
- The confusion matrix of the test results is shown in Fig. 13. As shown in Fig. 13(a), after authentication with the "apple" key, the accuracy of the test data on the suspicious model is 10%. As shown in Fig. 13(b), after authentication with the "rabbit" key, the accuracy of the test data on the suspicious model is 100%. Thus, *Bob* is the leaker of the DNN model.

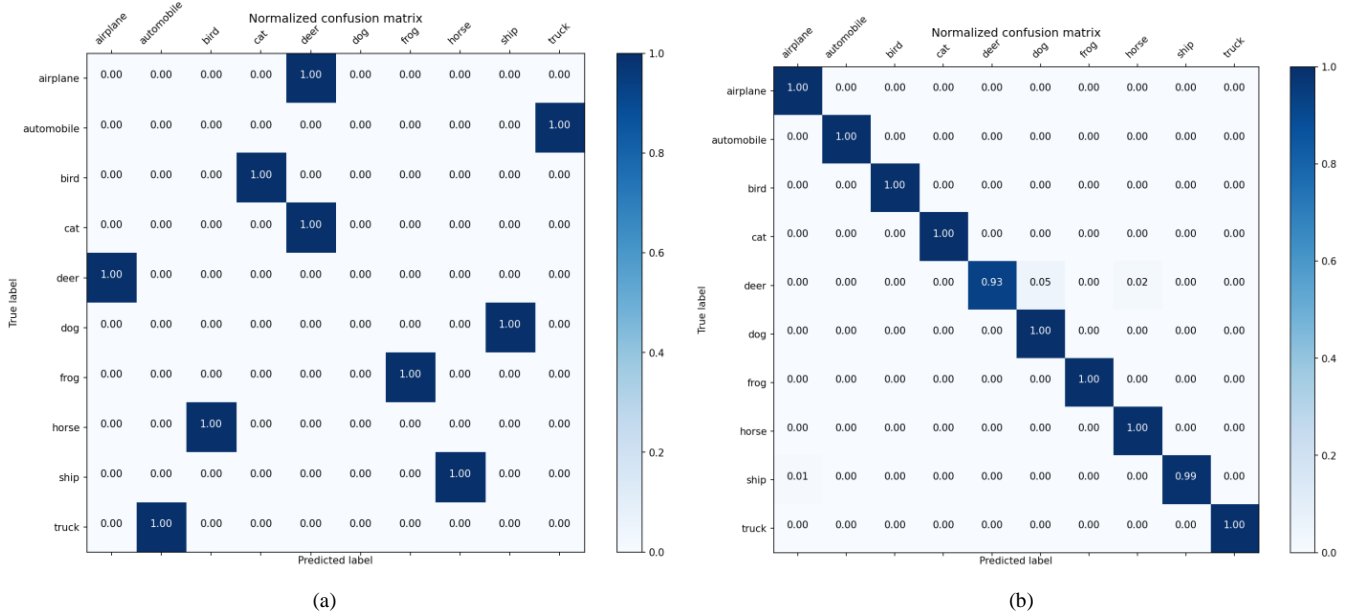

(a)   (b)

**Fig. 13.** Test results for suspicious models by selecting one image from each of the 10 CIFAR10 classes. Test result after authentication with (a) the "apple" key and (b) the "rabbit" key.

*E. ACPT extension*

As introduced in Section VI, the ACPT framework implements active copyright protection and traceability of the DNN model through the authorization control center and does not make changes to the DNN model itself. Therefore, the ACPT framework can be efficiently combined with existing excellent DNN watermarking works [7, 8, 15, 28, 29] to achieve double-layer protection of DNN models.

## VII. CONCLUSION

The traceability problem after the DNN model is stolen is a new problem driven by market demand for artificial intelligence. DNN watermarking and authorization control are two potential methods for addressing this problem. Herein, based on the idea of black-box neural network watermarking, combined with video framing and image perceptual hash algorithm, the PCPT framework was proposed using an additional class of DNN models. This improved the existing traceability mechanism and produced only a small number of false-positives. A DNN model ACPT framework was proposed based on the authorization control strategy and perceptual image hashing technology. Moreover, it used an authorization control center constructed by the detector and verifier, which realized stricter authorization control of the framework, established a strong connection between the user and model owner, and improved the framework security. The PCPT and ACPT frameworks addressed the traceability problem after the model is stolen by performing model copyright protection. Additionally, the generated key sample did not affect the quality of the original image but did support traceability verification. Experiments were conducted on the PCPT framework using 2 public datasets and 5 DNN models. The results showed that the PCPT framework dealt with the traceability problem after the model was stolen by utilizing the copyright protection of the DNN model. This approach was robust and reliable. Simultaneously, the VGG16, GoogLeNet, and ResNet18 models were taken as examples to demonstrate the effectiveness of the ACPT framework's authorization control and traceability performance. PCPT and ACPT frameworks are mainly used for image classification models. The proposed scheme still has limitations for the copyright protection of DNN models in the fields of image processing and natural language processing, and PCPT has the disadvantage of low efficiency when applied to a large number of users. Solving the above problems is the direction of our future research work.